# The conductance of the gated Aharonov - Bohm ring touching a quantum wire

I.A. Shelykh[1,2], N.G. Galkin[3] and N.T. Bagraev[4]

[1] *Physics and Astronomy School, University of Southampton, Highfield, Southampton, SO17 1BJ, UK*

[2] *St.Petersburg State Polytechnical University, Polytechnicheskaya 29, 195251, St. Petersburg, Russia*

[3] *Algodign LLC, Force Field Lab, 123379, Moscow, Russia*

[4] *A.F.Ioffe Physico- Technical Institute of RAS, 194021, St. Petersburg, Russia*

We analyse the conductance of the Aharonov - Bohm (AB) one- dimensional quantum ring touching a quantum wire. It is shown that in accordance with experimental data the period of the AB oscillations strongly depends on the chemical potential and the Rashba coupling parameter. The dependence of the conductance on the carrier's energy is shown to reveal the Fano resonances.

## 1. Introduction.

The mesoscopic physics became the intense research field in last two decades. The new intriguing phenomena were observed in this domain such as the quantum Hall effect [1], the conductance quantization in the quantum wires and quantum point contacts [2], and spontaneous spin polarization in quantum wires and quantum point contacts at low carrier concentration [3]. Many of these effects are of great interest from the fundamental point of view. Besides, the mesoscopic objects can serve as components of the electronic devices of new generation. The gated AB rings present a special interest as they can be used as basic components for the realization of the spin transistors [4], spin filters [5] and quantum splitters [6].

The configuration which is usually considered consists of the quantum ring with two symmetrically situated electrodes as it is shown at the inset of the Fig. 1. The conductance of such a structure depends both on the magnetic and electric fields applied perpendicular to the structure's interface. The former provides the Aharonov - Bohm phaseshift between the waves propagating in the clockwise and anticlockwise direction [7] thus resulting in the oscillations of the conductance [8]. The period of these oscillations is governed by the amplitude of the backscattering on the contacts between the ring and the leads [9]. If these contacts are almost transparent, the period of the oscillations is hc/e, while if the scattering on the contacts is strong and the probability of the carrier's round trips inside the ring is enhanced, the period reduces to the hc/2e and the

Aronov – Altshuler - Spivak (AAS) oscillations typical for the weak localization regime [10] are observed. The electric field applied perpendicular to the plane of the ring also affects the conductance. It has a double effect. Firstly, it shifts the subband's bottom inside the ring thus providing the change of the carrier's wavenumber. Therefore the conductance of the system can exhibit the oscillations in the complete analogy to those observed in the Fabry- Perot resonator. Secondly, it lifts the symmetry of the quantum well in the direction of the structure growth axis thereby inducing the Rashba spin - orbit coupling (SOI) inside the ring that is characterised by the SOI coupling parameter, $\alpha$. The latter depends linearly on the gate voltage [11] and creates the dynamical phaseshift between the waves propagating within the ring, which results in the Aharonov - Casher oscillations of the conductance [4,8].

In the present paper we analyze theoretically and experimentally the conductance of the structure shown in the Fig. 1. It consists of the quantum ring touching the quntum wire. Both the wire and the ring are considered to be narrow enough to support only one spin - degenerated propagating channel. The drain-source voltage causing the electric current in the system is taken to be weak enough, so the Landauer-Buttiker formula can be used for the calculation of the conductance at zero temperature [2]

$$G = \frac{e^2}{h}\left[T_\uparrow(\Phi,\alpha,k_F) + T_\downarrow(\Phi,\alpha,k_F)\right] \quad (1)$$

where $T_{\uparrow,\downarrow}(\Phi,\alpha,k_F)$ is the transmission coefficient for the electrons of the two opposite spin directions dependent on the magnetic flux, $\Phi$, Rashba coupling parameter, $\alpha$, and carrier's Fermi wavenumber, $k_F$. In the future analysis we suppose that the carrier's g- factor is small and the Zeeman splitting of the spin subbands can be neglected. Thus, the spin of the carrier is assumed to be affected only by the Rashba SOI. This condition is fullfiled in most realisations of the quantum AB rings [12]. The generalization for the case of the finite temperature reads [13]

$$G = \frac{e^2}{4\pi\hbar kT}\int_0^\infty \left[T_\uparrow(\Phi,\alpha,\varepsilon) + T_\downarrow(\Phi,\alpha,\varepsilon)\right]\cosh^{-2}\left(\frac{\varepsilon-\mu}{2kT}\right)d\varepsilon \quad (2)$$

where $\varepsilon$ is a carrier's energy. The conductance of the structure is thus determined by the transmission coefficient which can depend on the direction of the electron's spin if both the external magnetic field and the Rashba SOI are present (see below).

## 2. The calculation of the transmission amplitudes

In order to calculate the conductance of the system shown in the Fig. 1, it is necessary to denote the amplitudes of the propagating waves. For the commodity of the construction of the scattering matrix of the contact between the wire and the ring we introduced a short lead connecting them, the length of which is supposed to be equal to zero in the calculations presented below. In this configuration we have two contacts, 1 and 2, but each of them connects only three leads and thus can be characterised by the 3x3 scattering matrix, for which the parametrization is well known [14] and rather simple contrary to the 4x4 scattering matrix which should appear if the ring touching the wire without intermediate lead is considered.

Within the adiabatic approximation the spin of the carrier follows the direction of the effective magnetic field created by the Rashba SOI and the phase factors of the clockwise and anticlockwise propagating waves read

$$\tau_{1,2\uparrow} = \exp\left[i\left(2\pi k_\pm a \pm \frac{e\Phi}{\hbar c} \pm \theta_B\right)\right] \tag{3a}$$

$$\tau_{1,2\downarrow} = \exp\left[i\left(2\pi k_\mp a \pm \frac{e\Phi}{\hbar c} \mp \theta_B\right)\right] \tag{3b}$$

where the wavenumbers $k_\pm$ differ only if the Rashba SOI is taken into account, and in the adiabatic approximation read [6,15]

$$k_\pm = \pm\frac{m\alpha}{\hbar^2} + \sqrt{\left(\frac{m}{\hbar^2}\right)\left(\frac{m\alpha^2}{\hbar^2} + 2\varepsilon\right)} \quad, \tag{4}$$

$\theta_B$ notes the Berry phase, which can be calculated as (see Appendix 1)

$$\theta_B = 2\arctan\left(\frac{2\xi}{1-|\xi|^2}\right) \tag{5}$$

where $\xi = 2m\alpha a / \left(\hbar^2 + \sqrt{\hbar^4 + (2m\alpha a)^2}\right)$. Naturally, the Berry phase is zero, if the Rashba SOI is absent, while it reaches $\pi$ in the limit of the strong Rashba SOI when $2m\alpha / \hbar^2 \to \infty$ [4]. The spin index in the Eq. (3a), (3b) corresponds to the spin orientation as compare to the axis of the effective magnetic field created by the Rashba SOI.

The amplitudes of the waves are connected by the following set of the linear equations

$$\begin{pmatrix} B_{\uparrow,\downarrow} \\ b \\ A_{\uparrow,\downarrow} \end{pmatrix} = \begin{pmatrix} r_1 & \varepsilon_1 & t_1 \\ \varepsilon_1 & \sigma_1 & \varepsilon_1 \\ t_1 & \varepsilon_1 & r_1 \end{pmatrix} \begin{pmatrix} 1 \\ b\tau_{\uparrow,\downarrow} \\ 0 \end{pmatrix} \quad (6a)$$

$$\begin{pmatrix} f \\ b\tau_{\uparrow,\downarrow} \\ g \end{pmatrix} = \begin{pmatrix} r_2 & \varepsilon_2 & t_2 \\ \varepsilon_2 & \sigma_2 & \varepsilon_2 \\ t_2 & \varepsilon_2 & r_2 \end{pmatrix} \begin{pmatrix} f\tau_{1\uparrow,\downarrow} \\ b \\ g\tau_{2\uparrow,\downarrow} \end{pmatrix} \quad (6b)$$

where the 3x3 matrices appearing in these equations are nothing but unitary scattering matrices of the contacts 1 and 2. The physical meaning of the matrix elements as well as the relations between them are discussed in details in Refs 6,8,12 and are also given in the Appendix II. Using the Eq. (6a) one finds the transmission and reflection coefficients as follows

$$A_{\uparrow,\downarrow} = t_1 + \frac{\varepsilon_1^2 \tau_{\uparrow,\downarrow}}{1-\sigma_1 \tau_{\uparrow,\downarrow}} = \frac{t_1(1-\tau_{\uparrow,\downarrow})}{1-\sigma_1 \tau_{\uparrow,\downarrow}} \quad (7a)$$

$$B_{\uparrow,\downarrow} = r_1 + \frac{\varepsilon_1^2 \tau_{\uparrow,\downarrow}}{1-\sigma_1 \tau_{\uparrow,\downarrow}} = \frac{r_1(1+\tau_{\uparrow,\downarrow})}{1-\sigma_1 \tau_{\uparrow,\downarrow}} \quad (7b)$$

where we have used the relations between the elements of the scattering matrix given in the Appendix II. The factor $\tau$ gives the total phaseshift due to the passing through the AB ring depending on the properties of the contacts 1 and 2. The Eq. (6b) allows finding it in the form

$$\tau_{\uparrow,\downarrow} = \sigma_2 + \varepsilon_2^2 \frac{(\tau_{1\uparrow,\downarrow} + \tau_{2\uparrow,\downarrow}) + 2(t_2 - r_2)\tau_{1\uparrow,\downarrow}\tau_{2\uparrow,\downarrow}}{1 - r_2(\tau_{1\uparrow,\downarrow} + \tau_{2\uparrow,\downarrow}) + (r_2^2 - t_2^2)\tau_{1\uparrow,\downarrow}\tau_{2\uparrow,\downarrow}} \quad (8)$$

The formula (8) simplifies sufficiently in the case $\sigma_2 = 0$ which corresponds to the completely transparent contact 2. The further consideration will be limited by this case. One has

$$\tau_{\uparrow,\downarrow} = \frac{(\tau_{1\uparrow,\downarrow} + \tau_{2\uparrow,\downarrow})/2 - \tau_{1\uparrow,\downarrow}\tau_{2\uparrow,\downarrow}}{1 - (\tau_{1\uparrow,\downarrow} + \tau_{2\uparrow,\downarrow})/2} = \frac{\cos\left(\pm\frac{2\pi m\alpha a}{\hbar^2} + \frac{e\Phi}{hc} \pm \theta_B\right) - \exp\left(2\pi ia\sqrt{\left(\frac{m}{\hbar^2}\right)\left(\frac{m\alpha^2}{\hbar^2} + 2\varepsilon\right)}\right)}{\exp\left(-2\pi ia\sqrt{\left(\frac{m}{\hbar^2}\right)\left(\frac{m\alpha^2}{\hbar^2} + 2\varepsilon\right)}\right) - \cos\left(\pm\frac{2\pi m\alpha a}{\hbar^2} + \frac{e\Phi}{hc} \pm \theta_B\right)} \quad (9)$$

and the transmission amplitude is thus

$$A_{\uparrow,\downarrow} = 2t_1 \frac{\cos\left(\pm\frac{2\pi m\alpha a}{\hbar^2} + \frac{e\Phi}{hc} \pm \theta_B\right) - \cos\left(2\pi a\sqrt{\left(\frac{m}{\hbar^2}\right)\left(\frac{m\alpha^2}{\hbar^2} + 2\varepsilon\right)}\right)}{\left[\exp\left(-2\pi ia\sqrt{\left(\frac{m}{\hbar^2}\right)\left(\frac{m\alpha^2}{\hbar^2} + 2\varepsilon\right)}\right) + \sigma_1 \exp\left(2\pi ia\sqrt{\left(\frac{m}{\hbar^2}\right)\left(\frac{m\alpha^2}{\hbar^2} + 2\varepsilon\right)}\right)\right] - (1+\sigma_1)\cos\left(\pm\frac{2\pi m\alpha a}{\hbar^2} + \frac{e\Phi}{hc} \pm \theta_B\right)} \quad (10a)$$

$$B = 1 - A \quad (10b)$$

The parameter $\sigma_1$ characterizes the intensity of the coupling between the quantum ring and the wire and lies in the region [-1,1]. If $\sigma_1 = \pm 1$, the ring and the wire are completely uncoupled, and $T = |A|^2 \equiv 1$ which seems rather natural. For $\sigma_1 = 0$ the coupling is a maximum, and the transmission coefficients as a function of the external fields exhibit the oscillatory dependence.

## 3. Results and discussion

Firstly, the system in which the SOI is absent, $\alpha = 0$, is analyzed. In this case the transmission amplitude is independent of the spin direction because of the time inversion symmetry [16] and reads

$$A_\uparrow = A_\downarrow = 2t_1 \frac{\cos\left(\frac{e\Phi}{hc}\right) - \cos(2\pi ak)}{\left[\exp(-2\pi iak) + \sigma_1 \exp(2\pi iak)\right] - (1+\sigma_1)\cos\left(\frac{e\Phi}{hc}\right)} \quad (11)$$

where $k = \sqrt{2m\varepsilon/\hbar^2}$. The dependence of the conductance on the magnetic flux and the carrier's chemical potential is shown in the Figs. 2 and 3.

The dependence of the conductance on the magnetic field (Fig. 2) and the chemical potential (Fig. 3) has a resonant character. For zero temperature it turns to zero if the following condition is satisfied

$$e\Phi/hc \pm 2\pi ak = 2\pi n \quad (12)$$

Quite characteristically, the same equation determines the energetic spectrum of the isolated AB ring. The fact that the transmission falls to zero when the energy of the carrier corresponds to the energy of the bound level is by no means surprising being a typical feature when the interference of the bound state with energetic continuum takes place [17]. In the vicinity of the resonance the conductance should be described by the well-known Fano formula [18]. Indeed, decomposing the nominator and denominator of the Eq. (11) in the Taylor series in the vicinity of $k_0 = \pm \frac{1}{2\pi a}(e\Phi/hc + 2\pi n)$ and retaining only the terms linear in $k - k_0$ one has

$$T = T_0 \frac{(z+q)^2}{1+z^2} \quad (13)$$

where

$$z = \frac{\gamma}{\delta} 2\pi a(k - k_0) + \frac{\text{ctg}(2\pi ak_0)}{\gamma\delta} \quad (14)$$

with

$$\gamma^2 = \left[(1+\sigma_1)/(1-\sigma_1)\right]^2 + \text{ctg}^2(2\pi ak_0) \quad (14a),$$

$$\delta^2 = \left[(1+\sigma_1)/(1-\sigma_1)\right]^2 / \left(\left[(1+\sigma_1)/(1-\sigma_1)\right]^2 + \text{ctg}^2(2\pi a k_0)\right) \quad (14b),$$

$$T_0 = 4\left[\frac{r_1}{\gamma(1+\sigma_1)}\right]^2 \quad (14c)$$

The asymmetry factor q reads

$$q = \frac{\beta}{\alpha} = \frac{\sigma_1 + 1}{\sigma_1 - 1}\text{ctg}\left(\frac{e\Phi}{hc}\right) \quad (15)$$

The resonance is symmetric if q is either zero either infinity, which corresponds to the values of the flux $\frac{e\Phi}{hc} = \pi n$ ($q = \infty$) or $\frac{e\Phi}{hc} = \frac{\pi}{2} + \pi n$ ($q = 0$). In the first case the energy levels of the isolated AB ring are not split by the external magnetic field, while in the second case the splitting is exactly one half of the distance between the levels. Quite naturally, the width of the resonance given by the formula

$$\Gamma = \frac{(1+\sigma_1)(1-\sigma_1)}{2\pi a\left[(1+\sigma_1)^2 + (1-\sigma_1)^2 \text{ctg}^2(2\pi a k_0)\right]} \quad (16)$$

reduces to zero if $\sigma_1 = 0$, i.e. the wire and the ring are completely decoupled.

Secondly, the conductance as a function of the Rashba parameter, $\alpha$, is analyzed. If the external magnetic field is absent, $A_\uparrow = A_\downarrow$, and thus the transmitted and reflected currents are unpolarized. The conductance shown in the Fig. 4a is seen to exhibit irregular oscillations determined by the variations of $k_\pm$ due to the Aharonov - Casher effect and of $\theta_B$ with $\alpha$. The latter is only important when the SOI is relatively weak, $\xi = 2m\alpha a/\hbar^2 \sim 1$. The increase of $\alpha$ reduces $\theta_B$ up to $\pi$, as it is shown at the lower part of Fig.4a. The greater is the ring radius, the faster it moves to from 0 to $\pi$ with increasing the Rashba SOI parameter.

If the magnetic field and the Rashba SOI are present, the transmission amplitudes for the two spin components differ. It is clearly seen from Eq. (10a). Indeed, the transmission for the spin-up electrons turns into zero if

$$\frac{2\pi m\alpha a}{\hbar^2} + \frac{e\Phi}{hc} + \theta_B = 2\pi a\sqrt{\left(\frac{m}{\hbar^2}\right)\left(\frac{m\alpha^2}{\hbar^2} + 2\varepsilon\right)} + 2\pi n \quad (17)$$

while for the spin-down electrons the same condition is satisfied if

$$-\frac{2\pi m\alpha a}{\hbar^2} + \frac{e\Phi}{hc} - \theta_B = 2\pi a\sqrt{\left(\frac{m}{\hbar^2}\right)\left(\frac{m\alpha^2}{\hbar^2} + 2\varepsilon\right)} + 2\pi n \quad (17a)$$

The dependencies of the conductance on the Rashba coupling constant and $\Phi$ are shown at the Figs 4b, 4c. From Eq. (17) it follows that the transmitted and reflected currents now become spin - polarized, as it is shown in the insets of the corresponding figures. It is seen, that tuning of $\alpha$ or $\Phi$ allows the control of the spin polarization degree of the transmitted current, and thus the system considered can be used for spin filtering.

It is interesting to notice that the form of the AB oscillations in the magnetic field depends on the chemical potential and the Rashba SOI parameter, as it is clearly seen from Figs 2, 4c. In particular, these two quantities determine the relation between the amplitudes of the normal AB oscillations and the Aronov-Altshuler-Spivak oscillations of the half period. Contrary to the results of the works [19], the normal harmonics is present, except the points when $ka = \pi n/2$ (see Eq. (12)). This is easily understandable. Indeed, the presence of the half-period harmonics is provided by either interference of the two waves which made one round trip in the ring each moving in the opposite directions either by the interference of the wave which made two round trips with the wave which passed through the contact 1 without entering the ring. At the same time, the normal harmonics is provided by the interference of the wave which made one round trip in the ring with a free propagating wave. Clearly, the latter process can not be simply neglected. The relation between the powers of these two Fourier harmonics as a function of $\mu$ and $\alpha$ is shown in Figs. 5a and 5b. At some points the amplitude of the normal harmonics vanishes, which mean that the classical AAS oscillations [10] should be observed in this case. If the amplitudes of the two firs harmonics are zero, as it happens in Fig. 5a, the quarter-period oscillations should be observed. As both the chemical potential and the Rashba SOI parameter depend on the gate voltage, $V_g$ applied to the ring, these oscillations can be detected experimentally by varying its value. Indeed, the oscillations of the similar type were recently observed by Nitta and Koga [20].

## 4. Conclusions

In conclusion, we analyzed the conductance of the quantum ring touching a quantum wire as a function of the magnetic flux, the chemical potential and the Rashba SOI coupling parameter. The dependence of the conductance on the chemical potential was shown to be governed by the interference of the energetic continuum of the carriers inside the wire and the carrier's bound states inside the ring. The conductance in the resonance falls to zero and in its vicinity is well approximated by the Fano formula. The form of the AB oscillations in the system is determined by the chemical potential and the Rashba SOI parameter, which cause the oscillations in the relation to the intensities of the normal and half - period harmonics in the FFT. The dependence of the conductance on the Rashba SOI parameter, $\alpha$, is governed by both the geometrical Berry phase and the Aharonov-Casher phase. These conductance oscillations of the asymmetric nature caused by the contribution from the Aronov – Altshuler - Spivak oscillations seem to be revealed by the experimental studies of the one-dimensional semiconductor rings[20].

This work has been supported by SNSF in frameworks of the programme "Scientific Cooperation between Eastern Europe and Switzerland, Grant IB7320-110970/1.

**Appendix I. The calculation of the wavenumbers and the Berry phase**

According to the Ref. 14, the Hamiltonian of the AB ring with the Rashba SOI reads

$$H = \frac{\hbar^2}{2ma^2}\left\{\left[-i\frac{d}{d\varphi}+m\alpha a\left(\sigma_y \cos\varphi + \sigma_x \sin\varphi\right)\right]^2 - (m\alpha a)^2\right\} \quad \text{(AI-1)}$$

where we neglected the magnetic field as the AB phaseshift is already taken into account in Eq. (3). The eigenstates of the SOI Hamiltonian read

$$\psi_\uparrow^{(+)} = \frac{e^{+ik_+ a\varphi}}{\sqrt{1+\xi^2}}\begin{pmatrix} i\xi \\ e^{-i\varphi} \end{pmatrix};$$

$$\psi_\uparrow^{(-)} = \frac{e^{-ik_- a\varphi}}{\sqrt{1+\xi^2}}\begin{pmatrix} i\xi \\ e^{-i\varphi} \end{pmatrix};$$

$$\psi_\downarrow^{(+)} = \frac{e^{+ik_- a\varphi}}{\sqrt{1+\xi^2}}\begin{pmatrix} 1 \\ -i\xi \end{pmatrix};$$

$$\psi_\downarrow^{(-)} = \frac{e^{-ik_+ a\varphi}}{\sqrt{1+\xi^2}}\begin{pmatrix} 1 \\ -i\xi \end{pmatrix}.$$

(AI-2)

where $\xi = 2m\alpha a\big/\left(\hbar^2 + \sqrt{\hbar^4 + (2m\alpha a)^2}\right)$. The sign of the wavenumbers $k_{1,2}$ determines the direction of the motion, whereas their absolute values can be found from the following equation:

$$\left(\frac{2m\varepsilon}{\hbar^2} - k_\pm^2\right)\left[\frac{2m\varepsilon}{\hbar^2} - (k_\pm - 1/a)^2\right] - 4\frac{m^2\alpha^2}{\hbar^2}(k_\pm - 1/2a)^2 = 0 \quad \text{(AI-4)}$$

The eigenstates $\psi_1$ and $\psi_2$ correspond to the two orthogonal spin orientations. If the AB ring radius is large, so that $k_\pm \gg 1/a$, the spin of the two eigenstates is oriented in plane of the AB ring, towards or from the center. This result is easily understandable from the classical point of view. Indeed, neglecting the AB ring curvature, the effective magnetic field created by the Rashba SOI is given by the vector product of the external electric field and the carrier wavevector, $\mathbf{B}_{eff} = \frac{\alpha}{g_B \mu_B}[\mathbf{k} \times \mathbf{e}_z]$. In this case Eq. (AI-4) immediately reduces to the Eq. (4) in the text.

To determine the Berry phase in (3a)- (3b), let us remind that it can be calculated as a half of the matherial angle covered by the spin of the electron when it makes a round trip. The spin projection of the electron having the wavefunction (AI-2), (AI-3) makes with the z- axis an angle $\theta$ which can be calculated as

$$\tan\theta = \left|\frac{\sqrt{\langle S_x\rangle^2 + \langle S_y\rangle^2}}{\langle S_z\rangle}\right| = \frac{2\xi}{|\xi|^2 - 1} \tag{AI-5}$$

Quite importantly, it does not depend on $\varphi$ which allows a straightforward expression for the Berry phase, Eq. (5).

**Appendix II. Scattering matrices for the contacts 1 and 2**

Let us consider a contact connecting three leads 1, 2 and 3 as it is shown in Fig. 6. The scattering matrix of the contact connects the amplitudes of the incoming and reflected waves and reads

$$\begin{pmatrix} a_{out} \\ b_{out} \\ c_{out} \end{pmatrix} = \begin{pmatrix} r & \varepsilon & t \\ \varepsilon & \sigma & \varepsilon \\ t & \varepsilon & r \end{pmatrix} \begin{pmatrix} a_{in} \\ b_{in} \\ c_{in} \end{pmatrix} \tag{AII-1}$$

Here $r$ is a reflection amplitude of the leads 1 and 3, $t$ is a transmission amplitude from the lead 1 into lead 3 and from the leed 3 to the lead 1, $\varepsilon$ is the transmission amplitude from the leads 1 and 3 to the lead 2 and vice versa, $\sigma$ is a reflection amplitude of the lead 2. These parameters depend on the properties of the junction.

The number of independent matrix elements can be reduced, because the scattering matrix should be unitary owing to the conservation of the flux. Following Buttiker et al (See Refs. 13), the parameterisation of the scattering matrix reads:

$$r = \frac{\lambda_1 + \sigma}{2}, \tag{AII-2}$$

$$t = \frac{-\lambda_1 + \sigma}{2}, \tag{AII-3}$$

$$\varepsilon = \lambda_2\sqrt{\frac{1-\sigma^2}{2}} \tag{AII-4}$$

where $\lambda_{1,2}$ are either 1 or -1 (in the text both are taken to be +1). Therefore the effect of the QPCs on the scattering of the particle in the AB ring turns out to be defined by only one parameter, $\sigma$, which lies in the range $[-1,1]$ and characterises the coupling between the leads 1 and 3 and the lead 2. If $\sigma = 0$, this coupling is the strongest, while the $\sigma = \pm$ corresponds to the complete decoupling between the lead 2 and the leads 1,3.

**Figure captions**

**Fig. 1**. The amplitudes of the waves touching the quantum wire. The lead connecting the wire and the ring is introduced for the sake of the simplification of the calculations. Its length is then put equal to zero.

**Fig. 2** The dependence of the conductance on the magnetic flux for the different values of the temperature and the chemical potential if the Rashba SOI is absent. (a) – $\mu = 5 meV$, (b) – $\mu = 10 meV$, T=0 K and T=0.25 K. The radius of the ring is $a = 0.5$ mkm.

**Fig. 3** The dependence of the conductance on the chemical potential for the different values of the temperature and magnetic flux if the Rashba SOI is absent. (a) – $\Phi/\Phi_0 = 0$, (b) – $\Phi/\Phi_0 = \pi/4$, T=0 K and T=0.1 K. The radius of the ring is $a = 0.5$ mkm

**Fig. 4.**

(a). The dependence of the conductance on the Rashba SOI parameter for $\Phi/\Phi_0 = 0$. The lower part shows the Berry phase calculated by means of the formula (5). $\mu = 10 meV$, T=0 K and T=0.25 K. The radius of the ring is $a = 0.5$ mkm.

(b). The dependence of the conductance on the Rashba parameter for $\Phi/\Phi_0 = \pi/2$. The lower part shows the spin polarization degree of the transmitted current. $\mu = 10 meV$, T=0 K and T=0.25 K. The radius of the ring is $a = 0.5$ mkm.

(c). The dependence of the conductance on the magnetic flux for $\alpha = 8 \times 10^{-12}$ eVm, $\mu = 5 meV$. The radius of the ring is $a = 0.5$ mkm. The lower part shows the spin polarization degree of the transmitted current.

**Fig. 5.**

(a). The dependence of the intensity of the normal and half-period harmonics as a function of the chemical potential. $\alpha = 0$, $T = 0$, The radius of the ring is $a = 0.5$ mkm.

(b). The dependence of the intensity of the normal and half- period harmonics as a function of the Rashba SOI parameter $\alpha$. $\mu = 5$ meV, $T = 0$, The radius of the ring is $a = 0.5$ mkm.

**Fig. 6.** The amplitudes of the waves scattered from the conjunction of three 1D leads

**Fig. 1**

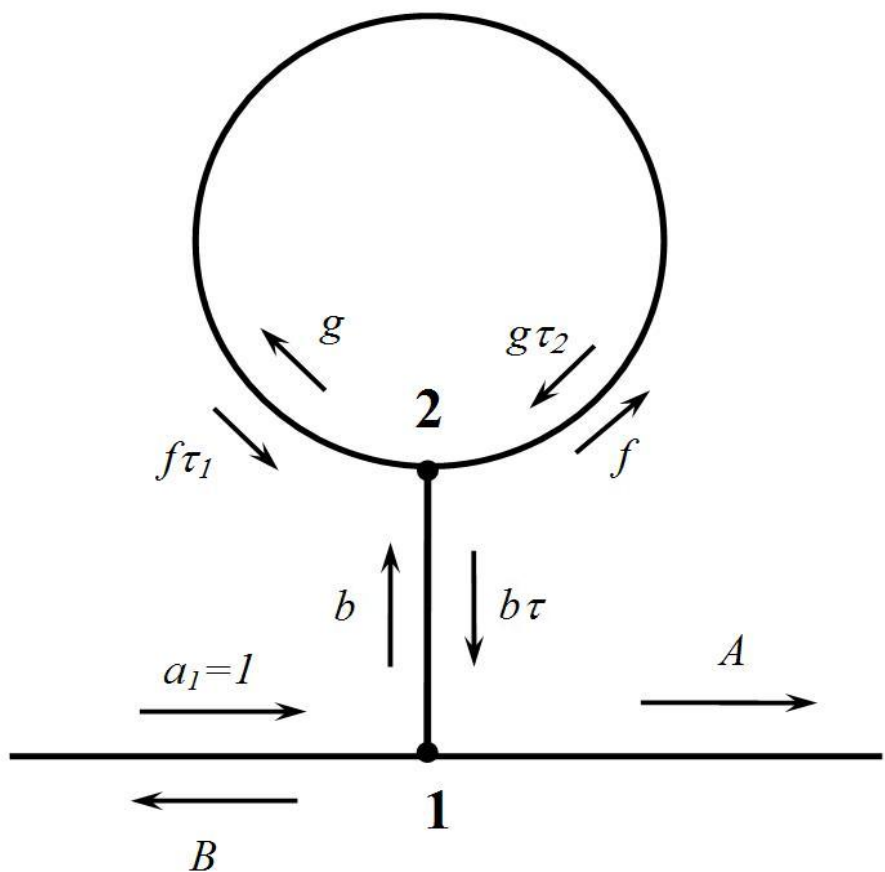

**Fig. 2**

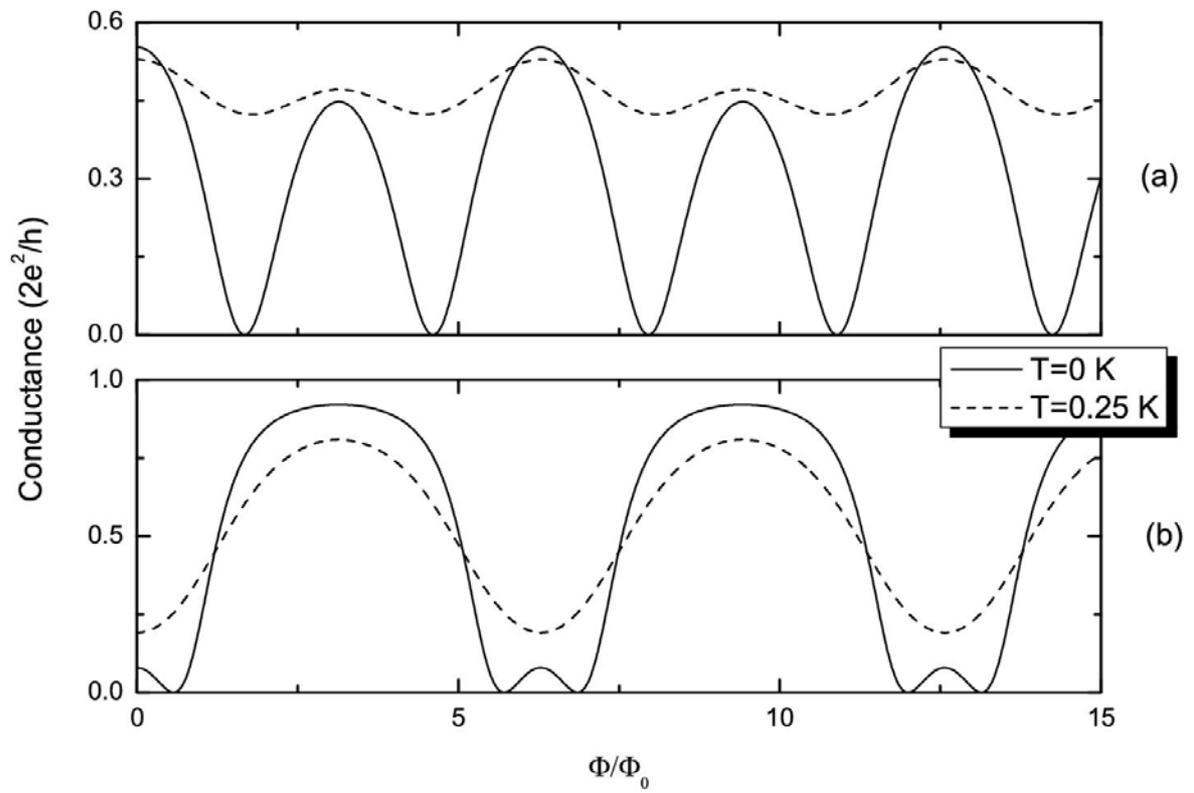

**Fig. 3**

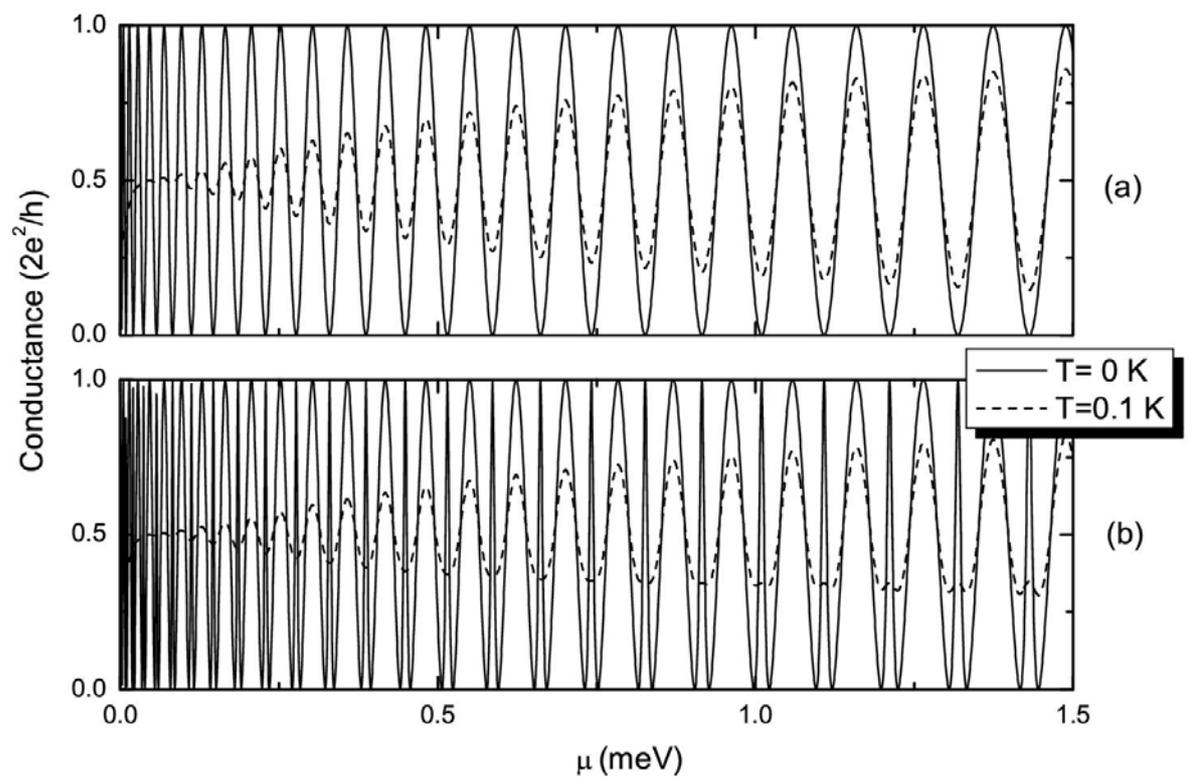

**Fig. 4a**

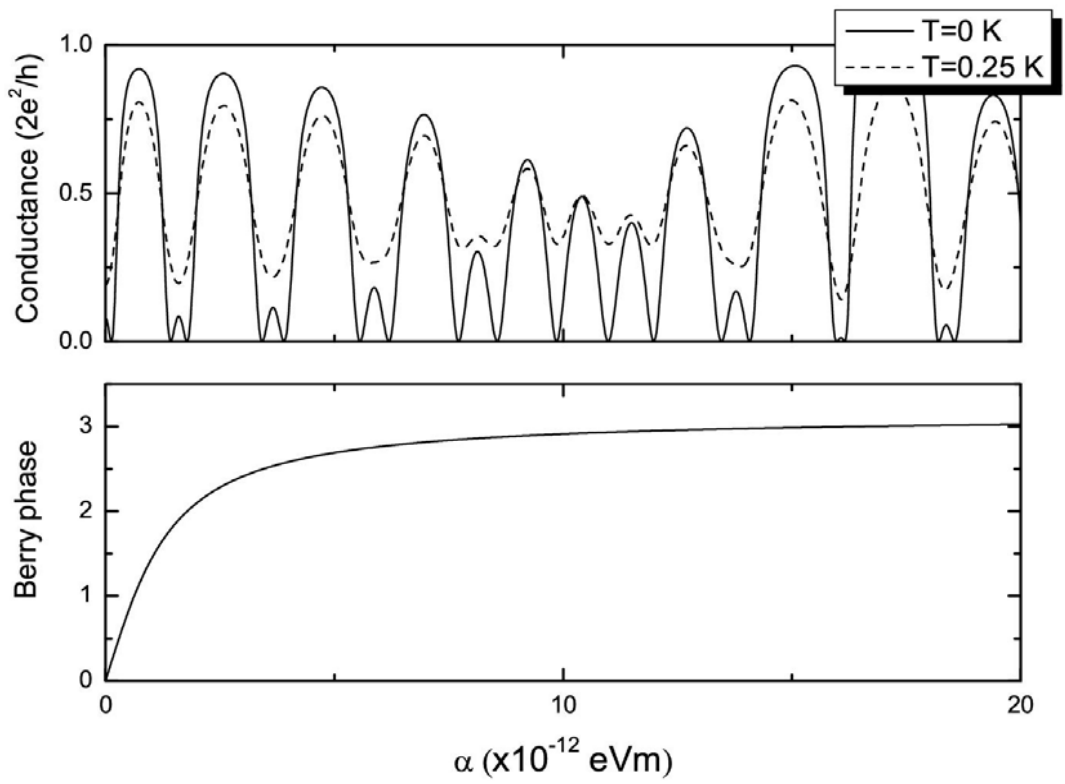

**Fig. 4b**

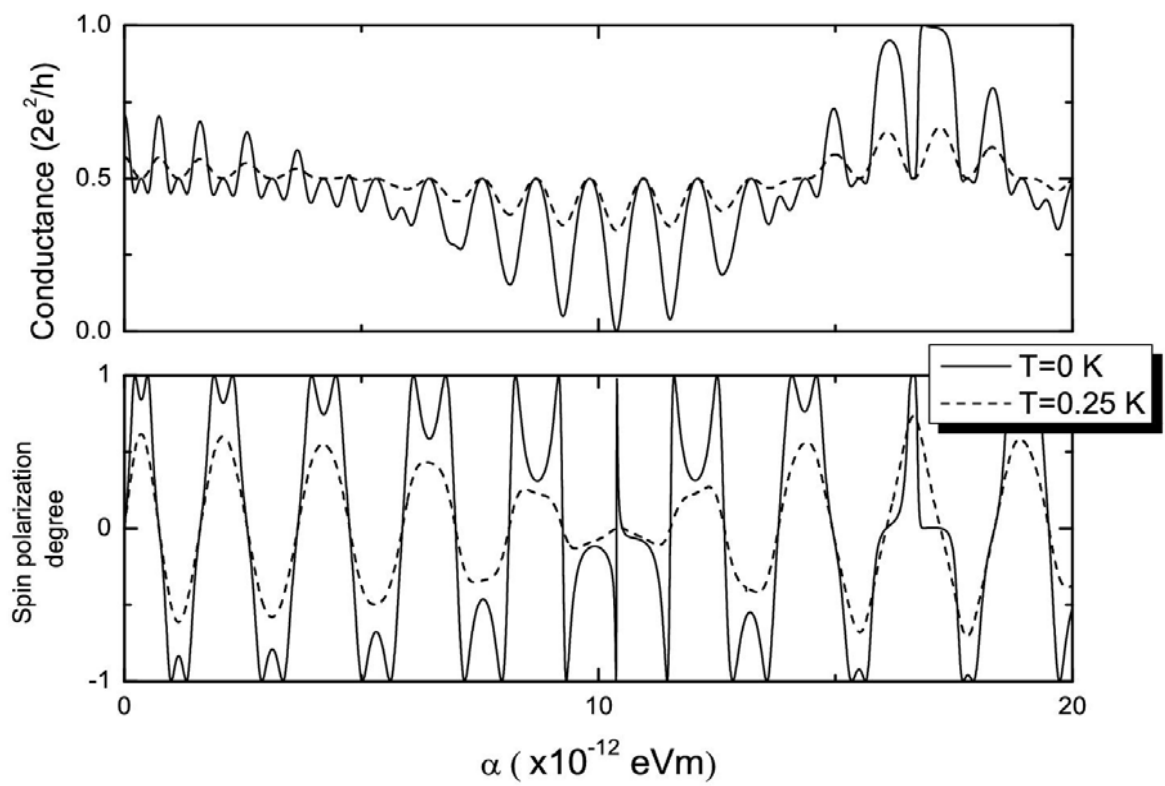

**Fig. 4c**

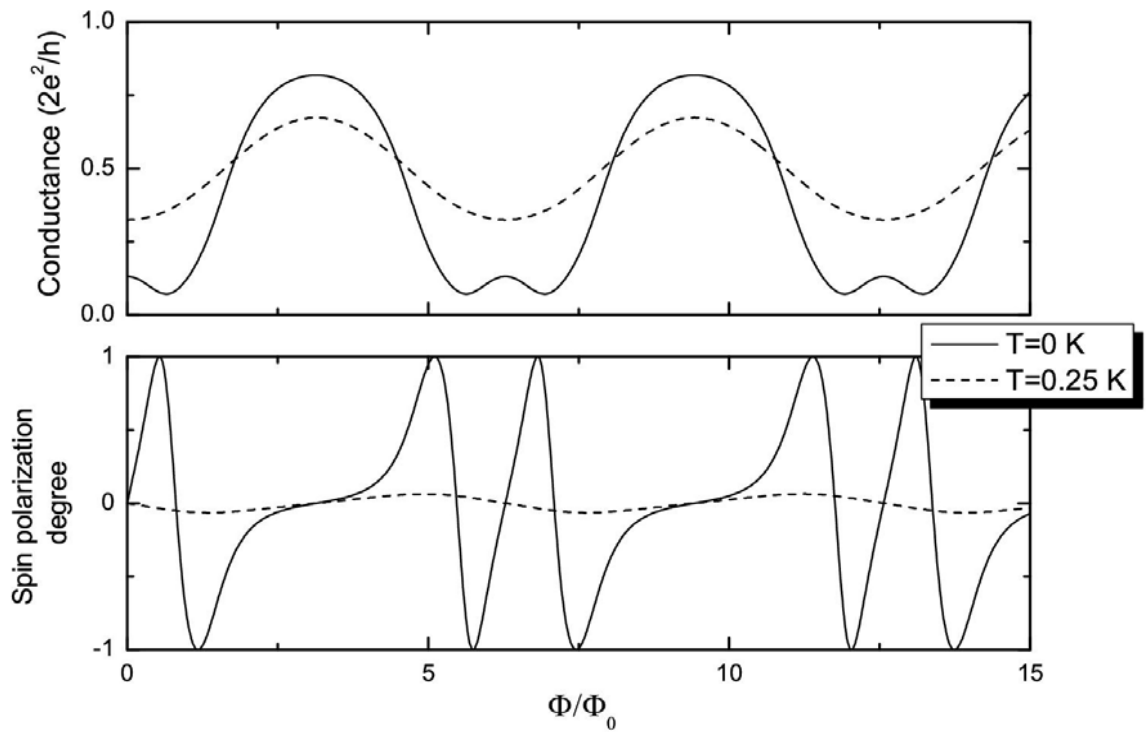

**Fig. 5a**

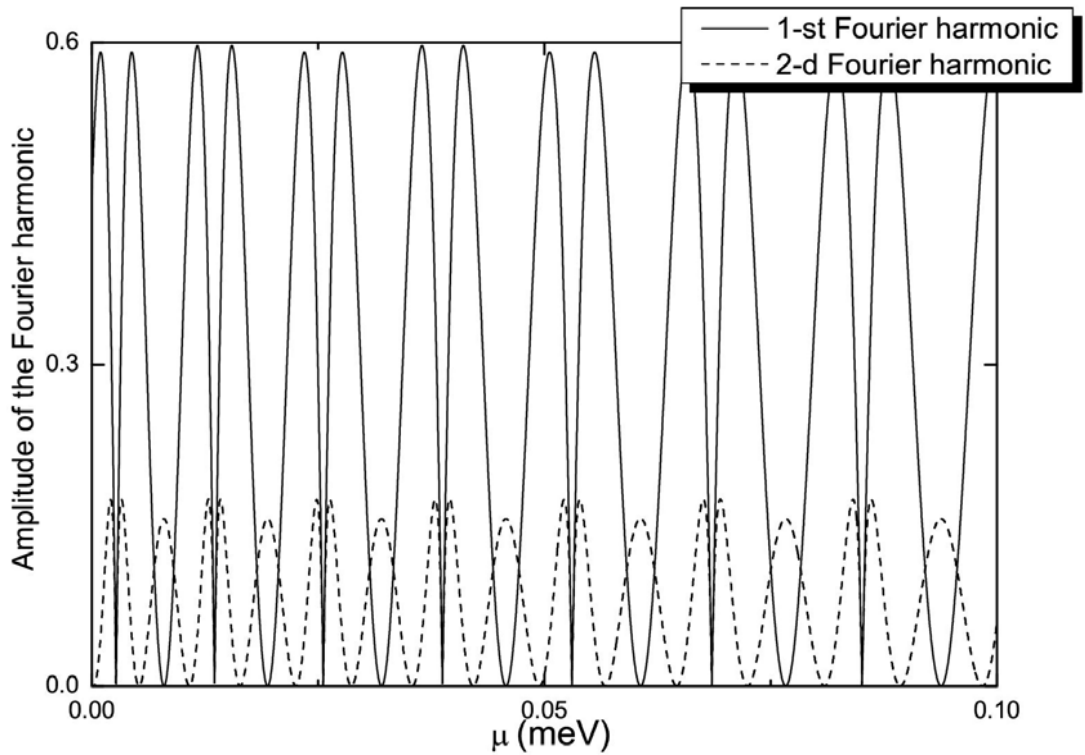

**Fig. 5b**

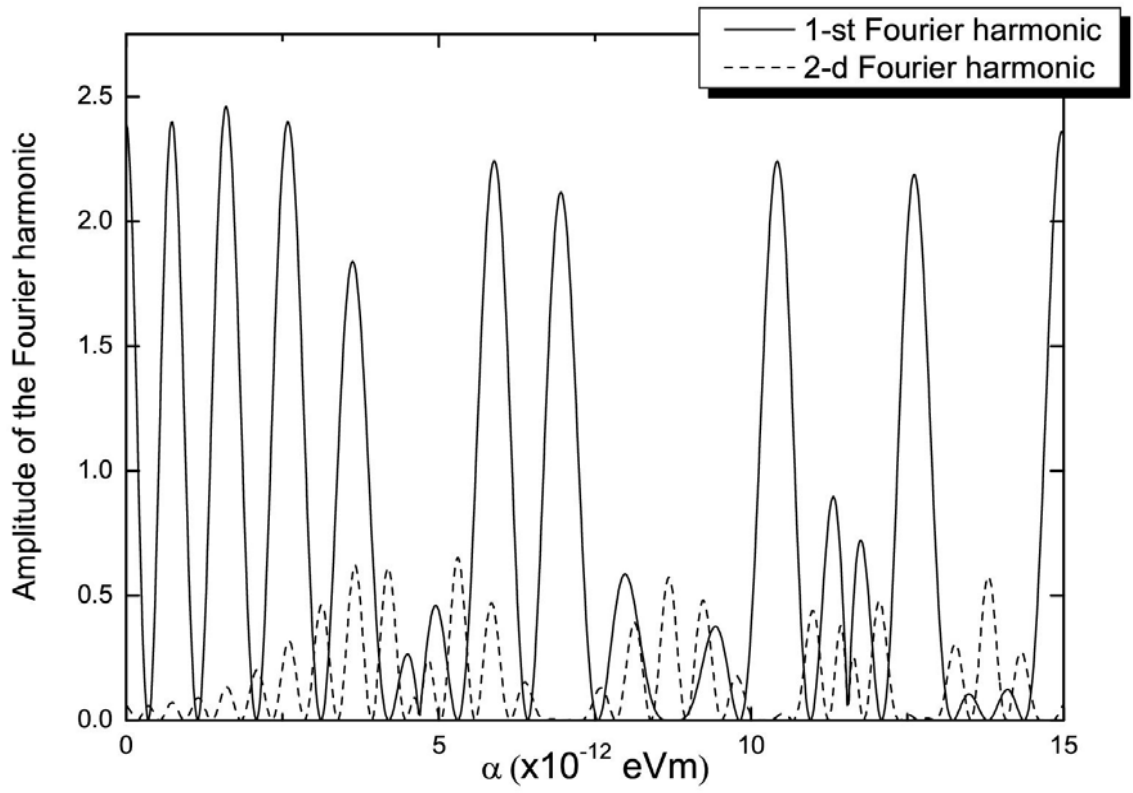

**Fig. 6**

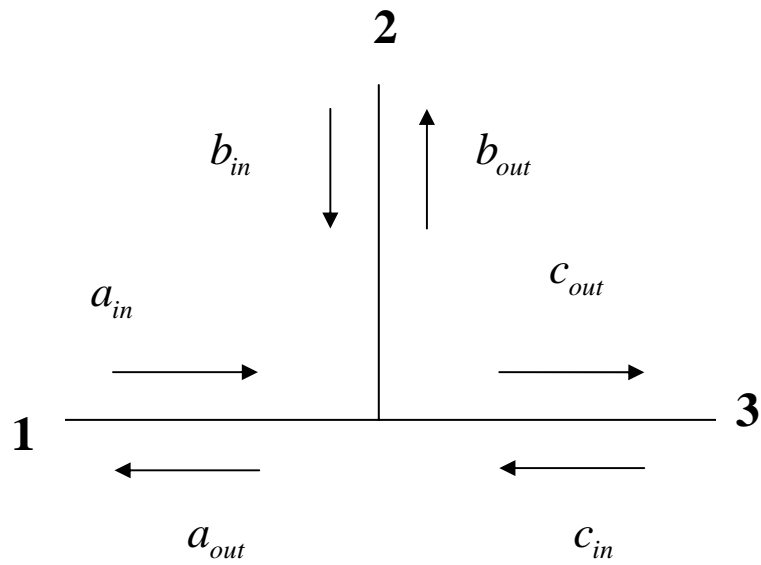